\documentclass[a4paper,11pt]{article}
\usepackage{pos}
\usepackage{xcolor}
\usepackage{wrapfig}
\usepackage[normalem]{ulem}

\renewcommand{\hookAfterAbstract}{%
\par\bigskip
\textsc{ArXiv ePrint}:
\href{https://arxiv.org/abs/2609.00230}{2609.00230}
}

\definecolor{NucleusFill}{RGB}{232,238,245}
\definecolor{NavyBlue}{RGB}{46,90,136}
\definecolor{RustR}{RGB}{181,70,30}

\newcommand{\orcid}[1]{\,\href{https://orcid.org/#1}{\includegraphics[width=9pt]{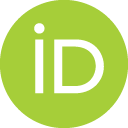}}\,}

\newcommand{\orcidFO}{0000-0001-6799-2436} %
\newcommand{\orcidJK}{0000-0002-3480-9823} %
\newcommand{\orcidBS}{0000-0003-3032-3305} %
\newcommand{\orcidRL}{0000-0003-4916-2898} %
\newcommand{\orcidPR}{0000-0002-8570-5506} %
\newcommand{\orcidBJS}{0000-0001-7908-1322} %

\newcommand{\smu}{Department of Physics, Southern Methodist University, Dallas, TX 75275-0175, USA}
\newcommand{\bnl}{Department of Physics, Brookhaven National Laboratory, Upton, NY 11973, USA}
\newcommand{\jlab}{Theory Center, Jefferson Lab, Newport News, VA 23606, U.S.A.}

\def\fhead#1{\noindent \textcolor{blue}{\hspace{-0pt} \textbf{{\small  \textbullet} 
{\textit{#1:}\quad}}}}

\title{Probing Dense Nuclear Matter at Small-x: \\
{\large A workflow for a global analysis framework}}
\ShortTitle{Probing Dense Nuclear Matter at Small-x}

\author[a]{Junaid~S.~Khan\orcid{\orcidJK}}
\author[a]{Rebecca~L.~Lustberg\orcid{\orcidRL}}
\author*[a]{Fredrick~Olness\orcid{\orcidFO}} %
\author[a,b]{Peter~Risse\orcid{\orcidPR}}
\author[c]{Bjoern~Schenke\orcid{\orcidBJS}}
\author[a]{Brandon~Stevenson\orcid{\orcidBS}}

\affiliation[a]{\smu}
\affiliation[b]{\jlab}
\affiliation[c]{\bnl}

\abstract{
The dipole model provides a powerful framework for describing high-energy nuclear interactions, particularly in the regime of dense gluonic matter. However, accurately evolving the dipole--nucleus scattering amplitude remains a major computational challenge because it is governed by nonlinear QCD evolution equations. To address this, we investigate a machine learning (ML) model as an efficient surrogate for the conventional numerical evolution. These ML-based approximations dramatically reduce the computational cost of global analyses while maintaining the accuracy required to describe a broad range of experimental data. We systematically evaluate the ML results for accuracy, computational efficiency, and ability to capture essential features of dipole evolution in nuclear environments.
These computational advancements will enable global analyses of diverse datasets within both the dipole and parton model frameworks, providing a more rigorous probe of nuclear structure in the dense regime.
Comparing both descriptions within a common fitting framework can provide precise constraints on the gluon distributions and advance our understanding of the quark and gluon structure of nuclei, particularly in the small-$x$ region.
}
\FullConference{XXXIII International Workshop on Deep Inelastic Scattering and Related Subjects (DIS2026)\\
4-8 May, 2026\\
Bologna, Italy\\}
\usepackage{lineno}
\modulolinenumbers[5]
\begin{document}
\maketitle

\def\figDipole{
\begin{wrapfigure}{r}{0.40\textwidth}
    \vspace{-30pt}
  \begin{center}
    \includegraphics[width=0.38\textwidth]{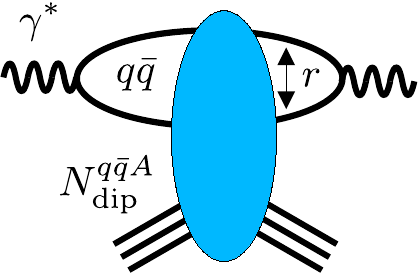}
    \vspace{-15pt}
  \end{center}
  \caption{Illustration of the dipole--nucleus scattering process.
  The virtual photon $\gamma^*$ splits into a $q \bar{q}$-pair with transverse separation $r$. 
  This color dipole system interacts with the nuclear target as described by the 
  {dipole nucleus cross section} $N_{\rm dip}^{q \bar{q} A}$.
  }
  \label{fig:dipole}
    \vspace{-15pt}
\end{wrapfigure}
}

\def\figBK{
\begin{figure}[t]
\vspace{-10pt}
    \centering
\vspace{-10pt}
    \includegraphics[width=1.0\linewidth]{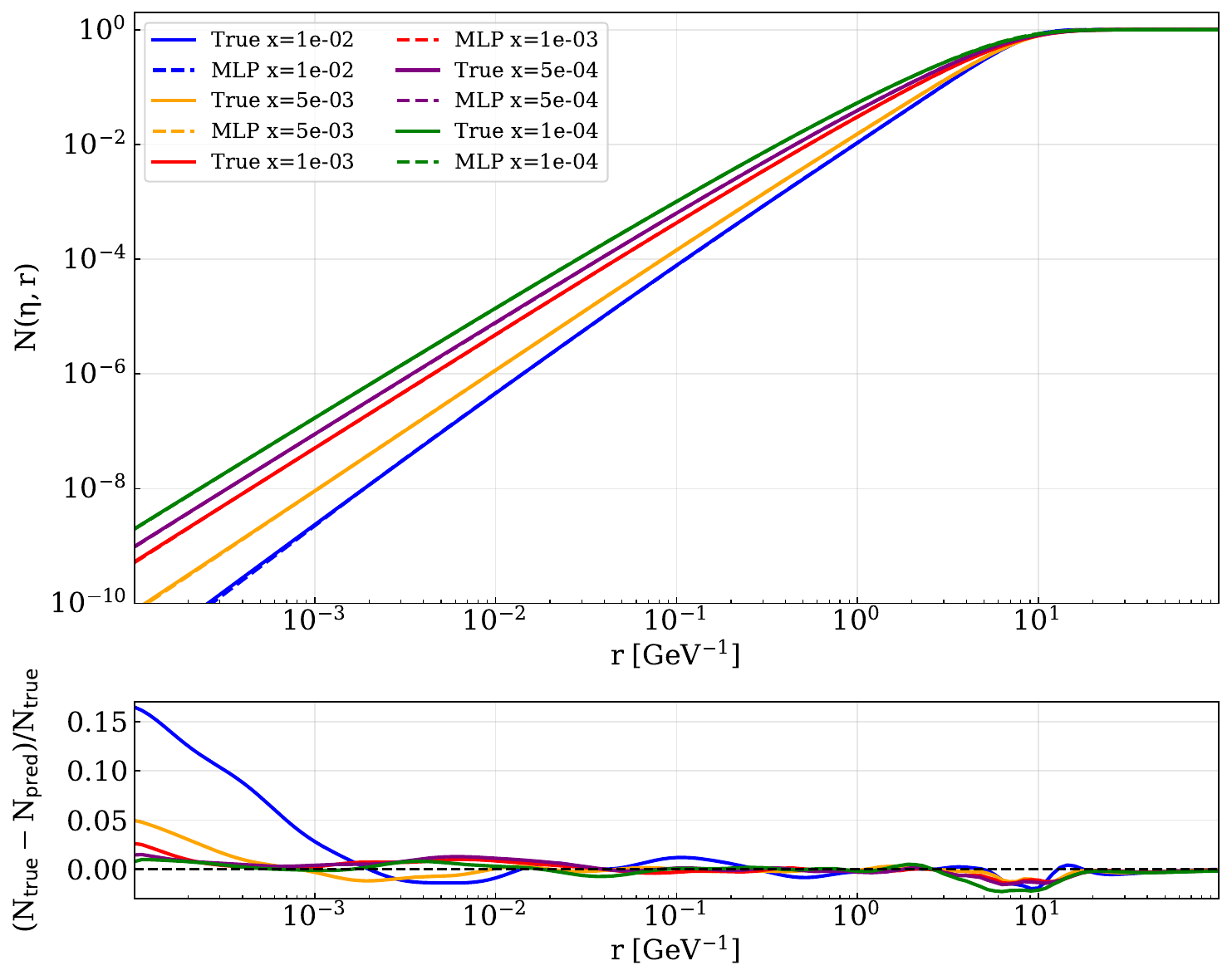}
    \caption{
Comparison of the MLP predictions (dashed lines) with the ``true'' BK evolved  solutions  (solid lines) for
selected $x$ values. 
The upper panel shows $N(\eta,r)$ as a   function of $r$,
and the lower panel shows the signed fractional error,
\mbox{$( N_{\mathrm{true}}-N_{\mathrm{pred}} )/N_{\mathrm{true}})$}.
The MLP closely reproduces the reference evolution over the full range of \(r\),
with the largest relative deviations occurring at very small \(r\), where
\(N\) is also extremely small.
} %
\vspace{-10pt}
    \label{fig:bkevl}
\end{figure}
}

\def\figHERA{
\begin{figure}[t]
\vspace{-10pt}
    \centering
\vspace{-10pt}
    \includegraphics[width=1.0\linewidth]{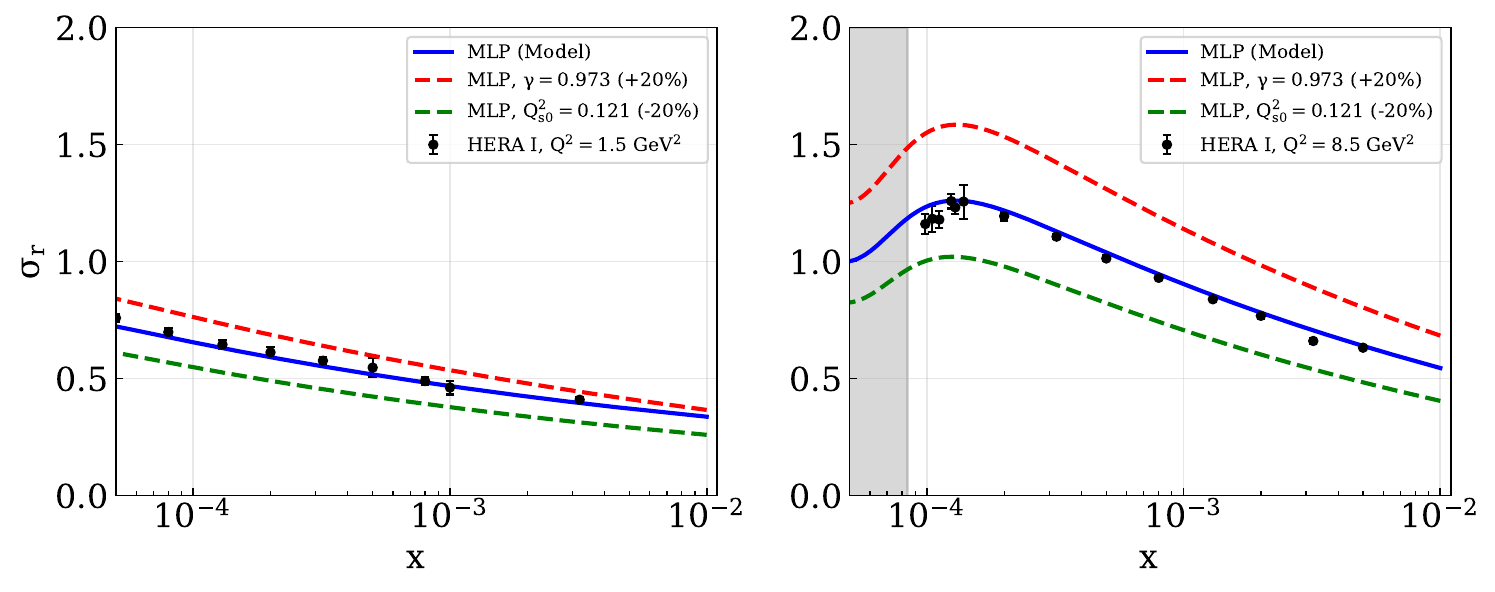}
    \caption{%
    Comparison of the MLP (Model) predictions with HERA reduced cross-section measurements~\cite{H1:2009pze}.
    The model parameters from Ref.~\cite{Lappi:2013zma} (blue solid curve) are
     \mbox{\(\gamma = 1.1350\)},
     \mbox{\(C^2 = 6.35\)},
     \mbox{\(Q_{s0}^2 = 0.165\)}, and
     \mbox{\(\sigma_0 = 84.5~\mathrm{GeV}^{-2}\)},
     and these agree well with the data. 
     The MLP allows us to easily vary the parameters as shown with the dashed curves.
    The shaded region indicates the non-physical region.
    }  %
    \label{fig:hera}
\end{figure}
}
\figDipole
\null\vspace{-30pt}
\fhead{Introduction}
Understanding the structure of hadrons and nuclei at high energies is a central goal of modern physics. While Quantum Chromodynamics (QCD) provides the fundamental theory of the strong interaction, two complementary QCD-based frameworks have proven especially effective for describing deep inelastic scattering and related processes that probe this structure.
The QCD parton model is highly successful in the dilute regime, where parton densities are low and nonlinear QCD effects are negligible. In contrast, the dipole model is tailored to the dense regime, where high gluon densities lead to collective nonlinear dynamics, including nuclear shadowing and gluon saturation, that lie beyond the conventional parton-model description.

\fhead{Dipole Model}
 In the dipole picture of Figure~\ref{fig:dipole}, a virtual photon first fluctuates into a color dipole, which subsequently scatters from a target hadron or nucleus. This framework naturally incorporates coherence effects and describes the transition from the dilute to the dense \mbox{small-x} regime, where increasing gluon densities drive nonlinear QCD dynamics, leading to gluon saturation and nuclear shadowing.
The physical cross section for the virtual photon ($\gamma^*$) nucleus ($A$) interaction process can be computed as follows:
\begin{equation}
  \sigma_{tot}^{\gamma^* A}(x,Q^2)
  = 
  \sigma_0 \, 
  \int \frac{d^2 \mathbf{r}}{4\pi} 
  \int_0^1 \!\! \frac{dz}{z(1-z)} 
    \left|\Psi^{\gamma^* \to q\bar{q}}(\mathbf{r},z)\right|^2
    \, N_{\mathrm{dip}}^{q\bar{q}A}(\eta,r) \quad .
\end{equation}
In this framework, the \textbf{light-cone wave function},  $\Psi^{\gamma^* \to q\bar{q}}$, gives the probability amplitude for a virtual photon to fluctuate into a quark--antiquark color dipole. The interaction of this dipole with the target nucleus is then encoded in the \textbf{dipole–nucleus cross section}, $N_{\mathrm{dip}}^{q\bar{q}A}$. 
The $\sigma_0$ is an effective transverse area parameter to
be fitted to data, and  $Q^2$ is the virtuality of the photon.
Here,  $x$ is the Bjorken-x variable; 
for convenience, we work with the logarithmic variable $\eta=\ln(x_0/x)$ 
where $x_0$ is a reference value. 

\fhead{Parton \& Dipole Model Comparison}
As we transition from the dilute to the dense partonic regime, the parton and dipole frameworks should smoothly connect in their common domain of validity. In this overlap region, both approaches are expected to produce consistent cross-section predictions. Once this correspondence is established, the dipole framework can be extended into the dense regime to quantify the impact of nonlinear QCD effects, such as gluon recombination and saturation, that lie beyond the conventional parton-model description.
This strategy was explored in Ref.~\cite{Armesto:2022mxy} and demonstrated to be an effective means of investigating the onset of nonlinear QCD dynamics. The comparison between the parton and dipole frameworks provides a quantitative benchmark for identifying the kinematic region where deviations from the linear parton-model description emerge, signaling the increasing importance of saturation effects.

\fhead{Global Analysis Framework}
Building on the comparison presented in Ref.~\cite{Armesto:2022mxy}, our objective is to develop a framework that enables direct and systematic comparisons between the parton and dipole formalisms. 
For example, the open-source \textbf{xFitter} platform may provide a natural foundation for integrating both approaches into a common global-analysis framework, enabling consistent comparisons of their predictions across a broad range of observables.
xFitter is an open-source analysis framework built on the QCD parton model, facilitating a wide range of phenomenological studies. Its broad applicability has led to its use in more than 100 scientific publications.~\cite{Alekhin:2014irh,xFitterDevelopersTeam:2018hym}.
The modular architecture of xFitter enables new processes and theoretical calculations to be incorporated seamlessly as external modules, as exemplified by the Fant\^omas framework \cite{Kotz:2025une}. In the same spirit, the conventional parton-model calculations can be replaced by a dipole-model module, allowing both approaches to be implemented and directly compared within a unified global analysis framework.

\fhead{Efficient Evaluation}
Phenomenological fits of the dipole–nucleus cross section to HERA data have provided valuable insights into the onset of gluon saturation and the interplay between linear and nonlinear QCD evolution. 
Although the ability to compute and compare the parton and dipole frameworks within a unified computational environment is highly attractive, performing such global analyses remains a significant computational challenge.
Since global analyses involve thousands of iterations over large datasets during parameter optimization, the repeated evaluation of dipole-model cross sections becomes a major computational bottleneck. Efficient and accurate computation of these observables is therefore essential.
We will examine the components required for this calculation in turn. 

\fhead{Light Cone Wave Function}
The light-cone wave function, $\Psi^{\gamma^* \to q\bar{q}}$, can be calculated perturbatively in QCD. At leading order, it is given by relatively simple combinations of Bessel functions that can be evaluated rapidly within each iteration of the fitting procedure. 
At higher orders, however, the light-cone wave function acquires substantially more complicated corrections, making its repeated numerical evaluation inside the fitting loop computationally prohibitive.
Fortunately, efficient strategies already exist for handling computationally intensive calculations. Techniques such as precomputed interpolation grids and k-factor methods enable fast approximations, making them well-suited for iterative global analyses. These methods have, for example, been implemented in xFitter.

\fhead{Dipole Nucleus Cross Section}
We will find that the fast evaluation of the dipole–nucleus cross section $N_{\mathrm{dip}}^{q\bar{q}A}\equiv N$ is a more challenging problem. 
To compute $N(\eta,r)$,  we take a model parametrization at an initial value $\eta_0$ and evolve this in $\eta$ using the \textbf{Balitsky--Kovchegov }(BK) equation. 
In the leading-logarithmic form, the BK equation is:~\cite{Balitsky:1995ub,Kovchegov:1999ua}
\begin{align}
    \frac{\partial N(\eta,r)}{\partial \eta} &=
    \int d^2 r_1 \, K(r,r_1,r_2) 
    \left[ 
    N(\eta,r_1) +N(\eta,r_2) -N(\eta,r) -N(\eta,r_1) N(\eta,r_2) 
    \right]
    \label{eq:bk}
\end{align}
with $r_2=r-r_1$ and $K$ is the evolution kernel. 
The first three terms in the integrand correspond to the linear \textbf{Balitsky–Fadin–Kuraev–Lipatov} (BFKL) evolution, while the quadratic term   encodes
the nonlinear saturation effects that tame the growth of gluon densities at small x.

\fhead{Initial Conditions}
 The initial condition for the evolution is typically parameterized at some starting value $\eta_0$ (defined by $x_0$) using a chosen model such as 
 Golec-Biernat-Wüsthoff (GBW) %
or 
McLerran-Venugopalan (MV$^\gamma$).
For the present analysis, we will use the 
MV$^\gamma$ model given by:~\cite{Albacete:2010sy, McLerran:1993ka,McLerran:1993ni}
\begin{equation}
    \label{eq:ic}
    N(\eta_0, r) = 1 - \exp\!\left[
      -\left(\frac{r^2 Q_{s0}^2}{4}\right)^{\!\gamma}
      \ln\!\left(\frac{1}{r\,\Lambda_{\mathrm{QCD}}} + e\right)
    \right]\, .
\end{equation}
Here, $Q_{s0}^2$ is the initial saturation scale, $\gamma$ controls the shape of the amplitude at small~$r$,
and 
$\Lambda_{QCD}$ is the fundamental scale parameter of QCD. 
In this work, we incorporate a running coupling into the BK evolution, resulting in running-coupling BK (rcBK) evolution, while retaining the McLerran-Venugopalan model (MV$^\gamma$) as the initial condition.
The details of this initial condition can be found in Ref.~\cite{McLerran:1993ka}, and this model includes an additional parameter, ``$C$'', which modifies the running of the coupling.

Given a choice of
model parameters $(Q_{s0}^2, \gamma, C, x_0)$, the MV$^\gamma$
initial condition together with rcBK evolution specifies the dipole–nucleus cross section $N(\eta_0,r)$ at the starting $\eta_0$. For any smaller Bjorken-$x$, corresponding to $\eta = \ln(x_0/x) > 0$, the evolved amplitude $N(\eta,r) \equiv N(x,r)$ is obtained by evolving the (rc)BK equation, Eq.~\eqref{eq:bk}, from $\eta_0$ to $\eta$.

\fhead{Machine Learning}
While numerical implementations of the BK equation exist, they are not computationally efficient enough for inclusion in a general-purpose global analysis framework that simultaneously describes a broad range of experimental data. This challenge becomes even more severe beyond leading order, where the added complexity of the BK evolution renders direct numerical solutions prohibitively expensive within an iterative fitting procedure.
To accelerate the evaluation of the dipole scattering
amplitude $N$ obtained from the BK  
evolution codes, we train a
feed-forward neural network, specifically a \textbf{multilayer perceptron} (MLP),
to emulate the numerical
solutions on a fixed $(\eta,r)$ grid. The resulting surrogate
model enables fast, continuous evaluation of the amplitude, which is particularly useful for fitting procedures
that require repeated calls to the BK evolution.

\fhead{Parameters}
Our goal is to approximate the solution of the BK evolution equation in the $(\eta,r)$ plane while explicitly retaining its dependence on the parameters that define the model parameters at $\eta_0$. For illustrative purposes, 
we will use the McLerran-Venugopalan model, MV${}^{\gamma}$,  for our initial conditions and allow the  \mbox{$\{\gamma, C, Q_{s0}\}$} parameters to vary.
Thus, we are working in a 5-dimensional space defined by the vector
$\overrightarrow{X} = \{\eta, r, \gamma, C, Q_{s0}\}$.
Here, $Q_{s0}$ is the initial saturation scale, 
$C$ is a parameter in the running coupling constant $\alpha_S$,
and $\gamma$ is an anomalous dimension used in the model; 
additional details of the  MV${}^{\gamma}$ can be found in Refs.~\cite{McLerran:1993ka,McLerran:1993ni}.

\fhead{MLP Architecture}
The MLP is implemented as a grid-wise feed-forward network operating on the full $(\eta, r)$ grid simultaneously. The input is a rank-5 tensor 
where each grid point is represented by $(\eta,   r, \gamma, C, Q_{s0})$.
The architecture consists of five fully connected hidden layers with widths $512, 512, 256, 256,$ and $128$, each followed by layer normalization,  Gaussian Error Linear Unit (GELU)  activation, and dropout regularization. A final linear layer produces pointwise predictions for $N(\eta,r)$.
This deep architecture captures the nonlinear dependence of the dipole amplitude on both kinematic variables and 
model parameters, while normalization and dropout improve training stability and generalization. Despite its capacity, the model remains computationally efficient due to weight sharing across the grid, enabling fast evaluation of the full $(\eta, r)$ dependence in a single forward pass.

\figBK

\fhead{Training Data}
To train the ML model, we generate a grid of dimension \mbox{$\{50,151,50,30,80\}$} in the variables \mbox{$\{\eta, r, \gamma, C, Q_{s0} \}$}.
Although generating the training dataset over the specified parameter ranges is computationally demanding, the calculation is naturally parallelizable and can therefore be completed efficiently on a CPU cluster within a few hours.
In contrast, this parallelization is applicable only to generating the training grid, where calculations for many independent parameter choices can be performed simultaneously. It does not provide the same benefit within the fitting loop, where BK evolution must be evaluated at arbitrary, continuously varying parameter points at each iteration of the global fit.

One might attempt to use the training grid directly by interpolating between the precomputed solutions during the fit. In practice, however, this strategy is impractical for two reasons. First, the training grid is exceptionally large, approximately 80 GB in the present example. Second, constructing a smooth interpolation in a high-dimensional parameter space requires many grid evaluations, making each fit iteration prohibitively expensive. Since the minimization procedure computes numerical derivatives of the $\chi^2$, the interpolation must also be sufficiently smooth and stable, further increasing the computational burden.

\fhead{Pros \& Cons}
The machine learning approach offers several compelling advantages. The trained model compresses an 80~GB training grid into a compact representation requiring only about 10~MB of weights, while reducing the evaluation time for each grid point to approximately 1~ms. In addition, it provides smooth, differentiable interpolations that are ideally suited for gradient-based $\chi^2$ minimization. These capabilities transform an otherwise computationally prohibitive calculation into one that can be efficiently incorporated into global analysis frameworks such as xFitter.

Conversely, the current approach requires the set of variables and their ranges to be specified in advance. As additional parameters are introduced, the size of the training grid increases accordingly. However, this growth may be mitigated by constructing a more judiciously chosen, sparser grid that retains sufficient coverage of the relevant parameter space.

\figHERA

\fhead{BK Results}
Figure~\ref{fig:bkevl} compares the dipole--nucleus cross section $N(\eta,r)$ from the approximate MLP with the computed BK evolution results across several $x$ values, demonstrating that the MLP model captures the overall evolution pattern across $r$. It also shows the fractional error as a function of $r$. Deviations are generally small ($\lesssim 5\%$) across the $r$ range with the exception of very small $r$ values where the absolute value of $N$ is also small. 
Furthermore, sample evaluations of the physical cross sections and structure functions find that the ML model yields results within $\sim 2\%$ or better across the kinematic range.

\fhead{HERA Comparisons}
In Figure~\ref{fig:hera}, we use the model parameters from Ref.~\cite{Lappi:2013zma} and demonstrate that our approach reproduces the HERA results, shown by the blue solid curve. Moreover, because variations in the model parameters are encoded in the ML weights, we can readily adjust these parameters within the trained network. This allows us to evaluate the model for arbitrary parameter variations around the reduced cross section $\sigma_r$, as illustrated by the dashed curves.

 \fhead{Conclusions}
In this work, we address this computational bottleneck by employing machine learning techniques to construct a fast, differentiable surrogate model for the dipole amplitude. Because the surrogate can evaluate the dipole amplitude orders of magnitude faster than directly solving the underlying evolution equations, it enables rapid exploration of the parameter space within an iterative fitting procedure. Once trained, the model provides an efficient means of evaluating the dependence of experimental observables on the underlying fit parameters.

By making BK evolution computationally tractable within a global analysis framework, this approach enables the simultaneous analysis of diverse datasets in both the dipole and parton-model formalisms. The resulting framework provides more stringent constraints on gluon distributions and advances our understanding of the structure of dense nuclear matter in the small-$x$ regime.

\fhead{Acknowledgments}
We gratefully acknowledge our SURGE collaborators, as well as 
Nestor Armesto, 
and Heikki Mäntysaari
for their insightful discussions.
We also thank our xFitter and nCTEQ collaborators for their valuable advice and support.
The work   was supported by the U.S.\ DoE, Office of Science, Office of Nuclear Physics, 
within the framework of the Saturated Glue (SURGE) Topical Theory Collaboration,
and by the U.S.\ DoE  Grant No.~DE-SC0010129.
The work of BS  is supported by the US Department of Energy Contract  No.~DE-SC0012704.
The work of FO was performed in part at the Aspen Center for Physics, which is supported by NSF grant PHY-2210452. 
The work of JK was supported by the O'Donnell Data Science and Research Computing Institute at SMU through a Graduate Research Fellowship. %
The work PR was supported by the US Department of Energy Contract No.~DE-AC05-06OR23177, under which Jefferson Science Associates, LLC operates Jefferson Lab.
 \null\vspace{-10pt}

\bibliographystyle{utphys}
\bibliography{refs}
\end{document}